\documentclass[10pt,journal]{IEEEtran}

\usepackage{psfrag}
\usepackage{epsfig}
\usepackage{graphicx}
\usepackage{multicol}
\usepackage{cite} 
\usepackage{amsthm}
\usepackage{subfigure}
\usepackage{nicefrac}

\RequirePackage{amsfonts}
\RequirePackage{amssymb}
\RequirePackage{amsopn}
\RequirePackage{amsmath}
\RequirePackage{pifont}
\RequirePackage{mathrsfs}
\RequirePackage{pstricks}

\def\atauxout{\csname @auxout\endcsname}%
\def\labelii#1{\immediate\write\atauxout{%
    \noexpand\newlabel{#1}{{\theenumii}{\thepage}}}}

\newcommand{\vect}[1]{\mathbf{#1}} 
\newcommand{\mat}[1]{\mathsf{#1}} 
 
\newcommand{\Reals}{\mathbb R}      
\newcommand{\Complex}{\mathbb C}    

\newcommand{\set}[1]{\mathcal{#1}}   

\newcommand{\const}[1]{\textnormal{\usefont{T1}{bc9}{r}{r}\selectfont #1}}
\newcommand{\Eb}{\mathcal{E}_{\textnormal{b}}} 
\newcommand{\Nzero}{\const{N}_{0}} 

\newcommand{\Hb}{H_{\textnormal{b}}}

\newcommand{\E}[2][]{\textnormal{\textsf{E}}_{#1}\!\left[#2\right]} 

\newcommand{\Normal}[2]{\mathcal{N}\!\left({#1},{#2}\right)} 

\renewcommand{\d}{\,\textnormal{d}}

\newcommand{\argmax}{\operatorname*{argmax}}

\newbox\measurebox %
\newlength{\firstmini} %
\newcommand{\mytextandeps}[2]{%
  \setbox\measurebox\hbox{\epsfig{file=#2}}
  \setlength{\firstmini}{\linewidth}%
  \addtolength{\firstmini}{-\wd\measurebox}%
  \addtolength{\firstmini}{-1em}%
  \begin{minipage}[t]{\firstmini}
    #1
  \end{minipage} \hfill %
  \setbox\measurebox\vbox{\unhbox\measurebox} %
  \setlength{\firstmini}{\ht\measurebox} %
  \addtolength{\firstmini}{\dp\measurebox} %
  \ht\measurebox=0pt \dp\measurebox=\firstmini %
  \box\measurebox
}%
  
\newgray{mygray}{0.9}
\makeatletter
  {\egroup     
  \noindent
  \centerline{
  \psshadowbox[shadowsize=0.3em,framesep=1.5em, fillstyle=solid, fillcolor=mygray]
              {\box\@tempboxa}
              }
  \par
  \vskip 0pt plus1ex 
  }%
\makeatother

\makeatletter
  {\egroup     
  \noindent
  \centerline{
  \psshadowbox[shadowsize=0.3em,framesep=1.5em, fillstyle=solid, fillcolor=white]
              {\box\@tempboxa}
              }
  \par
  \vskip 0pt plus1ex 
  }%

\newtheorem{theorem}{Theorem}
\newtheorem{lemma}{Lemma}
\newtheorem{corollary}{Corollary}
\newtheorem{proposition}{Proposition}

\newtheorem{condition}{Condition}
\newtheorem{example}{Example}

\title{Fundamental Limits of Communication with Low Probability of Detection}

\author{Ligong Wang,  Gregory W. Wornell,  and Lizhong Zheng}

\begin{document}

\maketitle 
\renewcommand{\thefootnote}{} 
\footnotetext[1]{This work was presented in part at the 2015 IEEE International Symposium of Information Theory (ISIT) in Hong Kong.

L. Wang is with ETIS (Equipes Traitement de l'Information et Syst\`emes), ENSEA, Universit\'e de Cergy-Pontoise, CNRS UMR 8051, France (e-mail: ligong.wang@ensea.fr). Part of this work was conducted while L.~Wang was with the Department of Electrical Engineering and Computer Science, and the Research Laboratory of Electronics, Massachusetts Institute of Technology, Cambridge, MA, USA.

G.W. Wornell and L. Zheng are with the Department of Electrical Engineering and Computer Science, and the Research Laboratory of Electronics, 
    Massachusetts Institute of Technology, Cambridge, MA, 
    USA (e-mail: gww@mit.edu; lizhong@mit.edu).

This work was supported in part by AFOSR under Grant
  No.~FA9550-11-1-0183, and by NSF under
  Grant No.~CCF-1319828.}
  
  
\renewcommand{\thefootnote}{\arabic{footnote}}
\begin{abstract}
This paper considers the problem of communication over a discrete memoryless channel (DMC) or an additive white Gaussian noise (AWGN) channel subject to the constraint that the probability that an adversary who observes the channel outputs can detect the communication is low. Specifically, the relative entropy between the output distributions when a codeword is transmitted and when no input is provided to the channel must be sufficiently small. For a DMC whose output distribution induced by the ``off'' input symbol is not a mixture of the output distributions induced by other input symbols, it is shown that the maximum amount of information that can be transmitted under this criterion scales like the square root of the blocklength. The same is true for the AWGN channel. Exact expressions for the scaling constant are also derived.
\end{abstract}


\section{Introduction}\label{sec:intro}
In many secret-communication applications, it is required not only that the adversary should not learn the content of the message being communicated, as in \cite{shannon49}, but also that it should not learn whether the legitimate parties are communicating at all or not. Such problems are often referred to as communication with \emph{low probability of detection (LPD)} or \emph{covert communication}. Depending on the application, they can be formulated in various ways.

In \cite{houkramer14} the authors consider a wiretap channel model \cite{wyner75}, and refer to this LPD requirement as \emph{stealth}. They show that stealth can be achieved without sacrificing communication rate or using an additional secret key. In their scheme, when not sending a message, the transmitter sends some random noise symbols to simulate the distribution of a codeword. There are many scenarios, however, where this cannot be done, because the transmitter must be switched off when not transmitting a message. Indeed, the criterion is often that the adversary should not be able to tell whether the transmitter is on or off, rather than whether it is sending anything meaningful or not. It is the former criterion that is considered in the current paper.

Our work is closely related to the recent works \cite{bashgoekeltowsley13,chebakshijaggi13,bloch16}. In \cite{bashgoekeltowsley13} the authors consider the problem of communication over an additive white Gaussian noise (AWGN) channel with the requirement that a wiretapper should not be able to tell with high confidence whether the transmitter is sending a codeword or the all-zero sequence. It is observed that the maximum amount of information that can be transmitted under this requirement scales like the \emph{square root} of the blocklength.\footnote{We adopt the usual terminology to use ``blocklength'' to refer to the total number of channel uses by a code. However, in the square-root case, the channel codes are not ``block codes'' in the traditional sense, because they cannot be used repeatedly. Indeed, repeated tramsmission would increase the eavesdropper's probability of detecting the communication.} In \cite{chebakshijaggi13} the authors consider a similar problem for the binary symmetric channel and show that the ``square-root law'' also holds. One major difference between \cite{bashgoekeltowsley13} and \cite{chebakshijaggi13} is that in the former the transmitter and the receiver use a secret key to generate their codebook, whereas in the latter no secret key is used. More recently, \cite{bloch16} studies the LPD problem from a resolvability perspective and improves upon \cite{bashgoekeltowsley13} in terms of secret-key length. 

In the current paper, we show that the square-root law holds for a broad class of discrete memoryless channels (DMCs).\footnote{The achievability part of the square-root law, but not the converse, is independently derived in \cite{bloch16}.} Furthermore, we provide exact characterizations for the scaling constant of the amount of information with respect to the square root of the blocklength for DMCs as well as AWGN channels, which is not done in \cite{bashgoekeltowsley13,chebakshijaggi13,bloch16}.

We do not assume that the eavesdropper observes a noisier channel than the intended receiver; instead, we assume that they both observe the same channel outputs. Our reason for dropping the wiretap structure is that, unlike in secret communication where the assumption that the eavesdropper observes a noisier channel allows one to obtain information-theoretic secrecy without using a secret key, in LPD problems the wiretap assumption does not bring essential new insights. In particular, the square-root law does not rely on the wiretap structure.\footnote{In fact, one can verify that the results in \cite{bashgoekeltowsley13} hold without the wiretap assumption; see Section~\ref{sec:AWGN} of the current paper for stronger results.} Hence, by putting the eavesdropper in the same position as the intended receiver, we allow ourselves to focus on the essence of the LPD-communication problem, while at the same time making our results more relevant in practice, the latter because in applications the legitimate parties usually cannot fully determine the statistical behavior of the eavesdropper's channel. We also note that extension of most of the results in the paper to wiretap channels is straightforward, part of which can be seen in \cite{bloch16}.


Because we do not assume a wiretap structure, contrary to \cite{chebakshijaggi13}, in our setting LPD communication is impossible without a secret key. We assume that such a key is available, and are not concerned with its length within the scope of this paper. 

We assume that the receiver does know when the transmitter is sending a message. This is a realistic assumption because the transmitter and the receiver can use part of their secret key to perform synchronization prior to transmission: They choose a (large enough) number of input sequences of a certain length such that each sequence induces an output distribution that is sufficiently different from the output distribution when there is no input to the channel, while on average these sequences induce an output distribution that is sufficiently close to the output distribution when there is no input. Using part of the secret key they randomly pick one of these sequences, which the transmitter sends to the receiver as a synchronization signal before sending a message.

One technical difference between \cite{bashgoekeltowsley13,chebakshijaggi13} and the present work is that the earlier works use total variation distance to measure probability of detection whereas we use \emph{relative entropy}, as \cite{houkramer14,cachin04}. Note that, when the relative entropy is given, the total variation distance can be upper-bounded using Pinsker's inequality \cite{csiszarkorner81}. See \cite{houkramer14} for further discussions on the relation between relative entropy and detectability. In practice, which of the two quantities is more relevant may depend on the actual application,\footnote{The total variation distance would be the right quantity to look at if one assumes equal probabilities for the transmitter sending and not sending a message, because it would correspond to the minimum probability of detection error by the eavesdropper. However, such an assumption is clearly unrealistic in practice.} whereas for theoretical analysis relative entropy is clearly easier to handle.

Summarizing the above discussions, we now briefly describe our setting:
\begin{itemize}
\item We consider a DMC whose input alphabet contains an ``off'' symbol. When the transmitter is switched off, it always sends this symbol.
\item The transmitter and the receiver share a secret key that is sufficiently long.
\item We assume that the adversary observes the same channel outputs as the intended receiver, i.e., there is no wiretap structure.
\item The LPD criterion is that the relative entropy between the output distributions when a codeword is transmitted and when the all-zero sequence is transmitted must be sufficiently small.
\end{itemize}

The square-root law has been observed in various scenarios in \emph{steganography} \cite{ker07,fridrich09,fillerfridrich09}. The setup in steganography that is most related to our work is as follows: a data file called the \emph{cover text} is generated according to some distribution, and a message must be concealed in this file subject to the constraint that the file should look almost unchanged. This is similar to the LPD setting in the sense that, when no message is to be conveyed, the encoder should not do anything, hence, in steganography the output is the original data file, whereas in LPD communications the output is pure noise. But steganography and LPD communications are essentially different: in steganography the data file is generated first and shown to the encoder, whereas in LPD communications noise is added to the codeword after the latter is chosen by the encoder. Hence the two types of problems require different analyses.

The rest of this paper is arranged as follows. In Section~\ref{sec:setup} we formulate the problem for DMCs and briefly analyze the case where the ``off'' input symbol induces an output distribution that can be written as a mixture of the other output distributions; the next two sections focus on the case where it cannot. In Section~\ref{sec:IXY} we derive formulas for characterizing the maximum amount of information that can be transmitted over any DMC under the LPD constraint. In Section~\ref{sec:var} we derive a simpler formula that is applicable to some DMCs. In Section~\ref{sec:AWGN} we formulate and solve the problem for AWGN channels. Finally, in Section~\ref{sec:conclusion} we conclude the paper with some remarks on future directions.

\section{Problem Formulation for DMCs}\label{sec:setup}
Consider a DMC of finite input and output alphabets $\set{X}$ and $\set{Y}$, and of transition law $W(\cdot|\cdot)$. Throughout this paper, we use the letter $P$ to denote input distributions on $\set{X}$ and the letter $Q$ to denote output distributions on $\set{Y}$. Let $0\in\set{X}$ be the ``off'' input symbol; i.e., when the transmitter is not sending a message, it always transmits $0$. Denote
\begin{equation}
	Q_0(\cdot) \triangleq W(\cdot|0).
\end{equation}
Without loss of generality, we assume that no two input symbols induce the same output distribution; in particular, $W(\cdot|x)=Q_0(\cdot)$ implies $x=0$.

	A (deterministic) code of blocklength $n$ for message set $\set{M}$ consists of an encoder $\set{M} \to \set{X}^n$, $m\mapsto x^n$ and a decoder $\set{Y}^n \to \set{M}$, $y^n\mapsto \hat{m}$. The transmitter and the receiver choose a \emph{random} code of blocklength $n$ for message set $\set{M}$ using a secret key shared between them. The adversary is assumed to know the distribution according to which the transmitter and the receiver choose the random code, but not their actual choice.\footnote{Note that we assume that the eavesdropper observes the same channel outputs as the intended receiver, so LPD communication is impossible with deterministic codes.}

The random code, together with a message $M$ uniformly drawn from $\set{M}$, induces a distribution $Q^n(\cdot)$ on $\set{Y}^n$. We require that, for some constant $\delta>0$,\footnote{All logarithms in this paper are natural. Accordingly, information is measured in nats.}
\begin{equation}\label{eq:LPD}
D\left(\left. Q^n \right\| Q_0^{\times n}\right) \le \delta.
\end{equation}
Here $Q_0^{\times n}$ denotes the $n$-fold product distribution of $Q_0$, i.e., the output distribution over $n$ channel uses when the transmitter is off. 

At this point, we observe that an input symbol $x$ with $\mathsf{supp}(W(\cdot|x)) \not\subseteq \mathsf{supp}(Q_0)$, where $\mathsf{supp}(\cdot)$ denotes the support of a distribution, should never be used by the transmitter. Indeed, using such an input symbol with nonzero probability would result in $D\left(\left. Q^n \right\| Q_0^{\times n}\right)$ being infinity. Hence we can drop all such input symbols, as well as all output symbols that do not lie in $\mathsf{supp} (Q_0)$, reducing the channel to one where
\begin{equation}\label{eq:suppQ0}
	\mathsf{supp} (Q_0) = \set{Y}.
\end{equation}
Throughout this paper we assume that \eqref{eq:suppQ0} is satisfied. Note that, for channels that cannot be reduced to one that satisfies \eqref{eq:suppQ0}, such as the binary erasure channel, nontrivial LPD communication is not possible. 

Our goal is to find the maximum possible value for $\log |\set{M}|$ for which a random codebook of length $n$ exists that satisfies condition \eqref{eq:LPD}, and whose average probability of error is at most $\epsilon$. (Later we shall require that $\epsilon$ be arbitrarily small.) We denote this maximum value by $K_n(\delta,\epsilon)$. 


We call an input symbol $x$ \emph{redundant} if $W(\cdot|x)$ can be written as a mixture of the other output distributions, i.e., if
\begin{equation}
	W(\cdot|x) \in \mathsf{conv} \left\{ W(\cdot|x')\colon x'\in\set{X}, x'\neq x\right\},
\end{equation}
where $\mathsf{conv}$ denotes the convex hull. As we shall show, $K_n(\delta,\epsilon)$ can increase either linearly with the blocklength $n$ or like $\sqrt{n}$, depending on whether $0$ is redundant or not.

\subsection{Case~1: input symbol $0$ is redundant}\label{sub:redundant}

This is the case where there exists some distribution $P$ on $\set{X}$ such that
\begin{subequations}\label{eq:degenerate}
\begin{IEEEeqnarray}{rCl}
	P(0) & = & 0 \label{eq:p00}\\
	\sum_{x\in\set{X}} P(x) W(\cdot |x) & = & Q_0(\cdot) . \label{eq:average}
\end{IEEEeqnarray}
\end{subequations}
In this case, a positive communication rate can be achieved:

\begin{proposition}\label{prp:linear}
If input symbol $0$ is redundant, then for any $\delta\ge 0$,
\begin{equation}
	\lim_{\epsilon\downarrow 0} \lim_{n\to\infty} \frac{K_n(\delta,\epsilon)}{n}= \max I(P,W),
\end{equation}
where the maximum is taken over input distribution $P$ that satisfies \eqref{eq:degenerate}.
\end{proposition}

\begin{IEEEproof}
First note that a random codebook generated IID according to $P$ that satisfies \eqref{eq:degenerate} yields $D(Q^n\|Q_0^{\times n}) = 0$. By the standard typicality argument~\cite{shannon48}, when the rate of the code is below $I(P,W)$, the probability of a decoding error can be made arbitrarily small as $n$ goes to infinity. Conversely, for a codebook whose empirical input distribution does not satisfy \eqref{eq:average}, $D(Q^n\|Q_0^{\times n})$ grows linearly in $n$ and is hence unbounded as $n$ goes to infinity. Finally, we check that any $P$ that does not satisfy \eqref{eq:p00} is suboptimal. Indeed, for any (nontrivial) $P$ that satisfies \eqref{eq:average} but not \eqref{eq:p00}, let $P'$ be $P$ conditional on $\set{X}\setminus \{0\}$, then $P'$ also satisfies \eqref{eq:average} and $I(P',W)>I(P,W)$.
\end{IEEEproof}

\begin{example}
Binary symmetric channel with an additional ``off'' symbol.
\end{example}

\begin{figure}[tbp]
\center
\vspace{-5mm}
\includegraphics[width=0.3\textwidth]{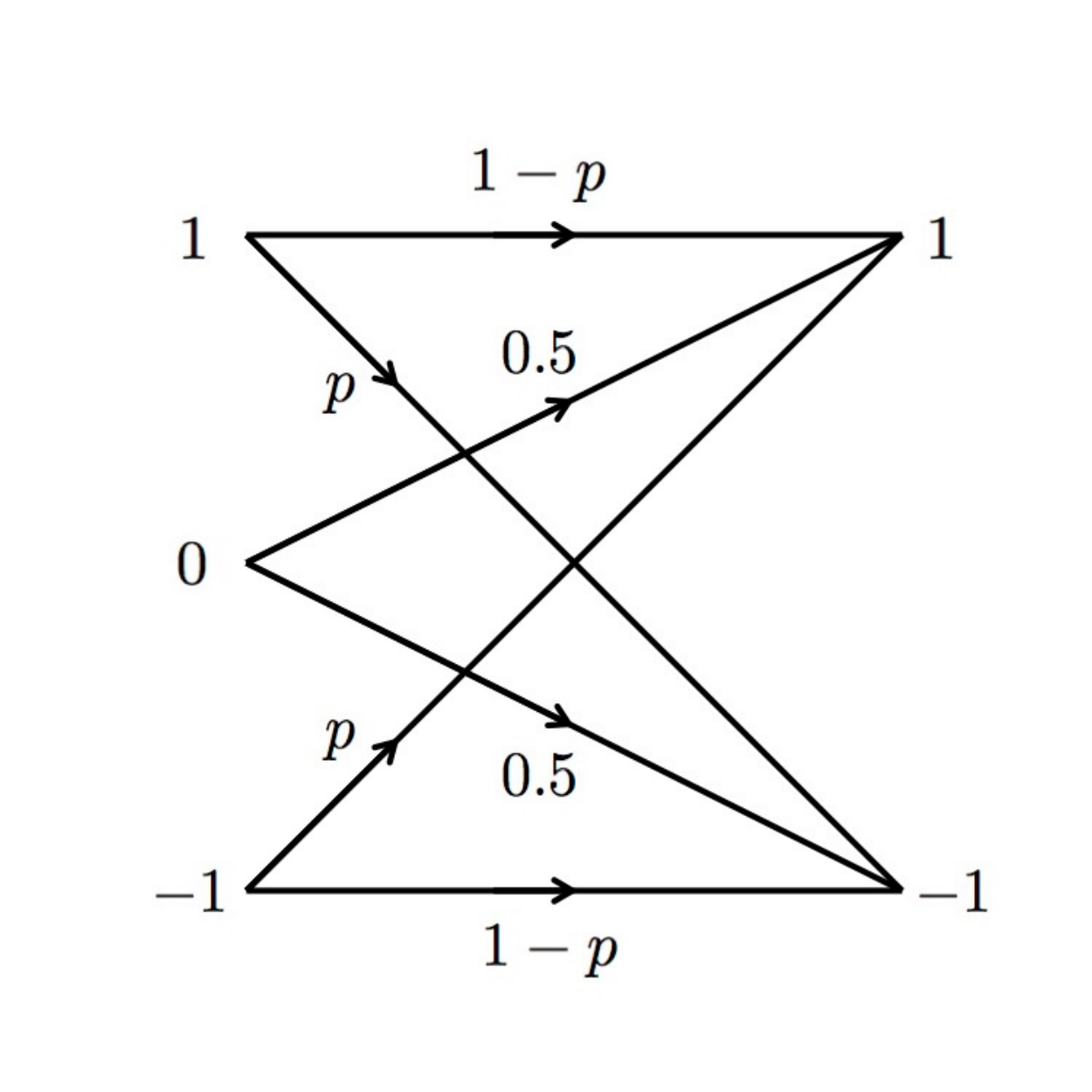}
\vspace{-5mm}
\caption{A binary symmetric channel on the alphabet $\{-1,1\}$ with cross-over probability $p$, with an additional ``off'' input symbol $0$ which induces a uniform output distribution. } \label{fig:BSC0}
\end{figure}

Consider a binary symmetric channel with an additional ``off'' symbol as shown in Fig.~\ref{fig:BSC0}. 
Its optimal input distribution for LPD communication is uniform on $\{-1,1\}$, and its capacity under the LPD constraint \eqref{eq:LPD} is the same as its capacity without this constraint, and equals $1-\Hb(p)$, where $\Hb(\cdot)$ is the binary entropy function.

\subsection{Case 2: input symbol $0$ is not redundant}\label{sub:notredundant}

\begin{figure}[tbp]
\center
\vspace{-3mm}
\includegraphics[width=0.3\textwidth]{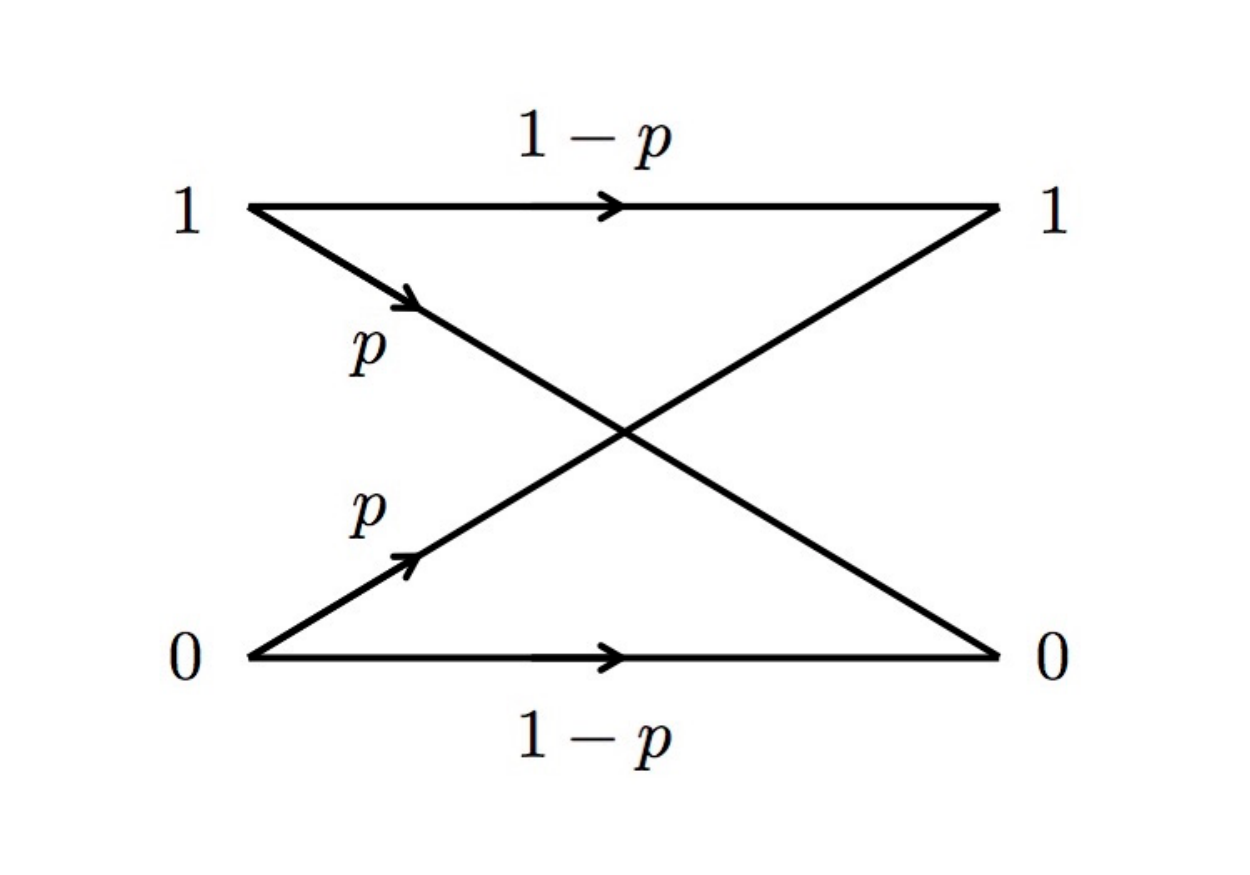}
\vspace{-5mm}
\caption{The binary symmetric channel with cross-over probability $p$. } \label{fig:BSC}
\end{figure}

This is the case where no $P$ satisfying \eqref{eq:degenerate} can be found. It is the focus of the next two sections. A simple example for this case is the binary symmetric channel in Fig.~\ref{fig:BSC}. 

We shall show that, in this case, $K_n$ grows like $\sqrt{n}$. Let
\begin{equation}\label{eq:defL}
	L \triangleq \lim_{\epsilon\downarrow 0} \varliminf_{n\to\infty} \frac{K_n(\delta,\epsilon)}{\sqrt{n\delta}},
\end{equation}
where $\varliminf$ denotes the limit inferior. 
Note that both $K_n(\delta,\epsilon)$ and $\delta$ have unit $\textnormal{nat}$, so $L$ has unit $\sqrt{\textnormal{nat}}$. We shall characterize $L$ in the next two sections. Note that, by definition, $L$ can be infinity, as it is in Case~1.

At this point, we provide some intuition why positive communication rates cannot be achieved in this case. To achieve a positive rate, a necessary condition is that a non-vanishing proportion of input symbols used in the codebook should be different from the ``off'' symbol $0$. This would mean that the average marginal distribution $\bar{P}$ on $\set{X}$ has a positive probability at values other than $0$ and, since $Q_0$ cannot be written as a mixture of output distributions produced by nonzero input symbols, the average output distribution $\bar{Q}$ must be different from $Q_0$ so $D(\bar{Q}\| Q_0)>0$.  This implies that $D(Q^n\|Q_0^{\times n})$ must grow without bound as $n$ tends to infinity, violating the LPD constraint \eqref{eq:LPD}.

\section{General Expressions for $L$ for All DMCs}\label{sec:IXY}
In this section we derive computable expressions for $L$. Our focus is on Case~2 where $0$ is not redundant, though some results also hold (in a trivial way) in Case~1 where $0$ is redundant. We first prove the following natural but nontrivial single-letter formula. 
\begin{theorem}\label{thm:IXY}
For any DMC,
\begin{equation}\label{eq:IXY}
	L =\max_{\{P_n\}} \varliminf_{n\to\infty} \sqrt{\frac{n}{\delta}} \,I(P_n,W)
\end{equation}
where the maximum is taken over sequences of joint distributions on $\set{X}\times \set{Y}$ induced by input distributions $P_n$ and channel $W$, whose marginals $Q_n$ on $\set{Y}$ satisfy
\begin{equation}\label{eq:deltan}
	D(Q_n\| Q_0) \le \frac{\delta}{n}.
\end{equation}
\end{theorem}

\emph{Remark:} Although the proof below does not guarantee that the limit inferior in \eqref{eq:IXY} can be replaced by the limit, this is indeed the case, as we show at the end of this section.

\begin{IEEEproof}[Proof of Theorem~\ref{thm:IXY}]
Proposition~\ref{prp:linear} shows that, when input symbol $0$ is redundant, $L=\infty$. This is consistent with Theorem~\ref{thm:IXY}. The rest of the proof focuses on Case~2 as in Section~\ref{sub:notredundant}, where $0$ is not redundant.

We first prove the converse part. This is done via Fano's inequality and manipulation of the information quantities. 

Suppose there exists a sequence of random codes satisfying \eqref{eq:LPD}, where, at blocklength $n$, the size of the codebook is $\exp(K_n)$, and the error probability is $\epsilon_n$ which tends to zero as $n$ tends to infinity. By a standard argument using Fano's inequality \cite{coverthomas91}, 
\begin{equation}
	K_n (1-\epsilon_n) -1 \le I(X^n;Y^n). \label{eq:singleletter10}
\end{equation}
Let $\bar{P}_n$ denote the average input distribution on $\set{X}$, averaged over the codebook and over the $n$ channel uses. We upper-bound $I(X^n;Y^n)$ in the usual way:
\begin{IEEEeqnarray*}{rCl}
	I(X^n;Y^n) & = & \sum_{i=1}^n I(X^n; Y_i|Y^{i-1})\\
	& = & \sum_{i=1}^n H(Y_i|Y^{i-1}) - H(Y_i|X^n, Y^{i-1})\\
	& = & \sum_{i=1}^n H(Y_i|Y^{i-1}) - H(Y_i|X_i)\\
	& \le & \sum_{i=1}^n I(X_i;Y_i)\\
	& \le & n I(\bar{P}_n,W), \IEEEyesnumber \label{eq:singleletter15}
\end{IEEEeqnarray*}
where the last step follows because, when the channel law is fixed, mutual information is concave in the input distribution. Combining \eqref{eq:defL}, \eqref{eq:singleletter10}, and \eqref{eq:singleletter15} yields
\begin{equation}
	L \le \varliminf_{n\to\infty} \sqrt{\frac{n}{\delta}}\, I(\bar{P}_n,W). \label{eq:single16}
\end{equation}

Next let $\bar{Q}_n$ denote the average output distribution on $\set{Y}$. Clearly, $\bar{Q}_n$ is the output distribution induced by $\bar{P}_n$ through $W$. Recall that $Q^n$ denotes the $n$-fold output distribution on $\set{Y}^n$. Further let $Q_{n,i}$ denote the marginal of $Q^n$ on the $i$th output $Y_i$. Let $Y^n$ have distribution $Q^n$, then (see also \cite{hou14})
\begin{IEEEeqnarray*}{rCl}
	D\left(\left. Q^n\right\|Q_0^{\times n} \right)
	& = & -H(Y^n) + \E[Q^n]{\log\frac{1}{Q_0^{\times n}(Y^n)}}\\
	& = & -\sum_{i=1}^n  H(Y_i|Y^{i-1}) + \E[Q^n]{\log\frac{1}{Q_0(Y_i)}}\\	
	& = & -\sum_{i=1}^n H(Y_i|Y^{i-1}) + \E[Q_{n,i}]{\log\frac{1}{Q_0(Y_i)}}\\
	& \ge & -\sum_{i=1}^n H(Y_i) + \E[Q_{n,i}]{\log\frac{1}{Q_0(Y_i)}}\\
	& = & \sum_{i=1}^n D(Q_{n,i}\|Q_0)\\
	& \ge & n D(\bar{Q}_n\|Q_0)\IEEEyesnumber
\end{IEEEeqnarray*}
where the last step follows because relative entropy is convex. This combined with \eqref{eq:LPD} implies that
\begin{equation}
	D(\bar{Q}_n\| Q_0) \le \frac{\delta}{n}.\label{eq:single23}
\end{equation}
Combining \eqref{eq:single16} and \eqref{eq:single23} proves the converse part of Theorem~\ref{thm:IXY}.

We next prove the achievability part. To this end, we randomly generate a codebook that satisfies \eqref{eq:LPD} and then show that, as the length of the codewords tends to infinity, the probability of a decoding error can be made arbitrarily small provided that the codebook has a size smaller than that determined by the right-hand side of \eqref{eq:IXY}. 

Let $\{P_n\}$ be a sequence of input distributions such that the induced output distributions $\{Q_n\}$ satisfy \eqref{eq:deltan}. For every $n$, we randomly generate a codebook by choosing the codewords IID according to $P_n$. The decoder performs joint-typicality decoding.

It is clear that the output distribution on $\set{Y}^{\times n}$ for this code is $Q^n = Q_n^{\times n}$ and that \eqref{eq:LPD} is satisfied. It remains to show that, provided that the size of the codebook is smaller than $\exp\big(nI(P_n,W) - \sqrt{n}\epsilon_n\big)$ for some $\epsilon_n$ tending to zero as $n$ tends to infinity, the probability of a decoding error can be made arbitrarily small. This cannot be shown using the asymptotic equipartition property \cite{shannon48}, or the information-spectrum method \cite{verduhan94,han03}, because we are in a situation where communication rate is zero. However, by slightly varying the methods in \cite{verduhan94,han03}, or using the one-shot achievability bounds as in \cite{wangcolbeckrenner09,polyanskiypoorverdu10}, we can obtain that the sequence $\{K_n\}$ is achievable provided
\begin{equation}\label{eq:Pliminf}
	\varliminf_{n\to\infty} \frac{K_n}{\sqrt{n}}\ge \textnormal{$P$-}\liminf_{n\to\infty} \frac{1}{\sqrt{n}} \log \frac{W(Y^n|X^n)}{Q_n^{\times n}(Y^n)},
\end{equation}
where $P$-$\liminf$ denotes the \emph{limit inferior in probability}, namely, the largest number such that the probability that the random variable in consideration is greater than this number tends to one as $n$ tends to infinity. Recalling \eqref{eq:defL}, to prove the achievability part of Theorem~\ref{thm:IXY}, it now suffices to show that the right-hand side of \eqref{eq:Pliminf} is lower-bounded by
$$ \varliminf_{n\to\infty} \sqrt{n} I(P_n,W).$$
We show a slightly stronger result which is
\begin{equation}\label{eq:inprobability}
	\frac{1}{\sqrt{n}} \log\frac{W(Y^n|X^n)}{Q_n^{\times n}(Y^n)} - \sqrt{n}\, I(P_n,W) \to 0 \quad \textnormal{in probability}
\end{equation}
as $n$ tends to infinity.
To this end, first note 
\begin{equation}
	\E{\frac{1}{\sqrt{n}} \log \frac{W(Y^n|X^n)}{Q_n^{\times n}(Y^n)}} = \frac{1}{\sqrt{n}} I(X^n; Y^n) =  \sqrt{n} \, I(P_n, W).
\end{equation}
It then follows by Chebyshev's inequality that, for any constant $a>0$,
\begin{IEEEeqnarray}{rCl}
	\lefteqn{\mathsf{Pr} \left[ \left| \frac{1}{\sqrt{n}} \log\frac{W(Y^n|X^n)}{Q_n^{\times n}(Y^n)} - \sqrt{n} \, I(P_n,W) \right| \ge a \right]}~~~~~~~~~~~~~~~~~~~~~\nonumber\\
	& \le & \frac{1}{a^2} \mathsf{var} \left( \frac{1}{\sqrt{n}} \log \frac{W(Y^n|X^n)}{Q_n^{\times n}(Y^n)} \right).\IEEEeqnarraynumspace
\end{IEEEeqnarray}
Thus, to prove \eqref{eq:inprobability}, it suffices to show
\begin{equation} \label{eq:tozero}
	\mathsf{var} \left( \frac{1}{\sqrt{n}} \log \frac{W(Y^n|X^n)}{Q_n^{\times n}(Y^n)} \right) \to 0
\end{equation}
as $n$ tends to infinity. To show \eqref{eq:tozero}, we first simplify this variance to
\begin{IEEEeqnarray}{rCl}
	\mathsf{var} \left( \frac{1}{\sqrt{n}} \log \frac{W(Y^n|X^n)}{Q_n^{\times n}(Y^n)} \right)
	& = & \frac{1}{n} \sum_{i=1}^n \mathsf{var} \left( \log\frac{W(Y_i|X_i)}{Q_n(Y_i)} \right) \nonumber \\
	& = & \mathsf{var}\left( \log\frac{W(Y|X)}{Q_n(Y)}\right). \label{eq:var29}
\end{IEEEeqnarray}
The variance on the right-hand side of \eqref{eq:var29} is upper-bounded by the second moment:
\begin{IEEEeqnarray}{rCl}
	\lefteqn{\mathsf{var}\left( \log\frac{W(Y|X)}{Q_n(Y)}\right)}~~~~~~~~\nonumber\\
	& \le & \E[P_n\circ W]{\left( \log\frac{W(Y|X)}{Q_n(Y)}\right)^2} \nonumber\\
	& = & P_n(0) \, \E[Q_0]{\left( \log\frac{Q_0(Y)}{Q_n(Y)}\right)^2} \nonumber\\
	& & {} + \sum_{x\neq 0} P_n(x) \,\E[W(\cdot|x)]{\left(\log\frac{W(Y|x)}{Q_n(Y)}\right)^2}.\label{eq:var30}\IEEEeqnarraynumspace
\end{IEEEeqnarray}
Here we use $P_n\circ W$ to denote the joint distribution on $\set{X}\times\set{Y}$ induced by input distribution $P_n$ through channel $W$.
To prove \eqref{eq:tozero}, it suffices to show that both terms on the right-hand side of \eqref{eq:var30} tend to zero as $n$ tends to infinity. For the first term, note that \eqref{eq:deltan} requires that 
\begin{equation}\label{eq:QntoQ0}
Q_n\to Q_0
\end{equation}
as $n$ tends to infinity, so
\begin{equation}
	\lim_{n\to\infty} \log\frac{Q_0(y)}{Q_n(y)} = 0, \quad \forall y\in\set{Y},
\end{equation}
which further implies (recall that $|\set{Y}|$ is finite so one can switch the order of limit and expectation)
\begin{equation}
\lim_{n\to\infty} \E[Q_0]{\left( \log\frac{Q_0(Y)}{Q_n(Y)}\right)^2} = 0.
\end{equation}
Thus, since $P_n(0)$ is bounded between $0$ and~$1$, the first term on the right-hand side of \eqref{eq:var30} tends to zero as $n$ tends to infinity. To analyze the second term on the right-hand side of \eqref{eq:var30}, recall our assumption that $Q_0$ cannot be written as a mixture of the other output distributions. Thus, to have \eqref{eq:QntoQ0} we need
\begin{equation}\label{eq:Pntoone}
	\lim_{n\to\infty} P_n(0) = 1,
\end{equation}
so
\begin{equation}\label{eq:Pntozero}
	\lim_{n\to\infty} P_n(x)=0, \quad \forall x\neq 0.
\end{equation}
We next use \eqref{eq:QntoQ0} to obtain (recall again that $|\set{Y}|$ is finite)
\begin{IEEEeqnarray}{rCl}
\lefteqn{\lim_{n\to\infty} \E[W(\cdot|x)]{\left(\log\frac{W(Y|x)}{Q_n(Y)}\right)^2}}~~~~~~\nonumber \\
& = & \E[W(\cdot|x)]{\left(\log\frac{W(Y|x)}{Q_0(Y)}\right)^2},
\end{IEEEeqnarray}
which is finite for every $x\in\set{X}$, $x\neq 0$, because $Q_0(y)>0$ for every $y\in\set{Y}$; recall \eqref{eq:suppQ0}. This combined with \eqref{eq:Pntozero} implies that the second term on the right-hand side of \eqref{eq:var30} tends to zero as $n$ tends to infinity.

We have now established that the right-hand side of \eqref{eq:var30} tends to zero as $n$ tends to infinity, which further establishes \eqref{eq:tozero} and, hence, \eqref{eq:inprobability}. This concludes the achievability part of Theorem~\ref{thm:IXY}.
\end{IEEEproof}

Using Theorem~\ref{thm:IXY} we derive the following computable expression for $L$.

\begin{theorem}\label{thm:general}
	For any DMC satisfying \eqref{eq:suppQ0},  whose ``off'' input symbol $0$ is not redundant, and which has at least one input symbol other than $0$,\footnote{By our assumption, this input symbol induces an output distribution that is different from $Q_0$, so the channel is not trivial.} $L$ is positive and finite, and is given by
\begin{equation}\label{eq:general}
	L = \max_{\tilde{P}\colon \tilde{P}(0)=0} \frac{ \sum_{x\in\set{X}} \tilde{P}(x) D\left( \left.(W(\cdot|x) \right\| Q_0\right)}{\sqrt{\displaystyle \frac{1}{2}\sum_{y\in\set{Y}} \frac{\big(\tilde{Q}(y) - Q_0(y)\big)^2}{Q_0(y)}}},
\end{equation}
where $\tilde{Q}$ is the output distribution induced by $\tilde{P}$ through $W$. 
\end{theorem}

Before proving Theorem~\ref{thm:general} we note that, for some channels, such as the next example, \eqref{eq:general} is very easy to compute. 
\begin{example}\label{ex:BSC}
Binary symmetric channel.
\end{example}

\begin{figure}[tbp]
\center
\includegraphics[width=0.4\textwidth]{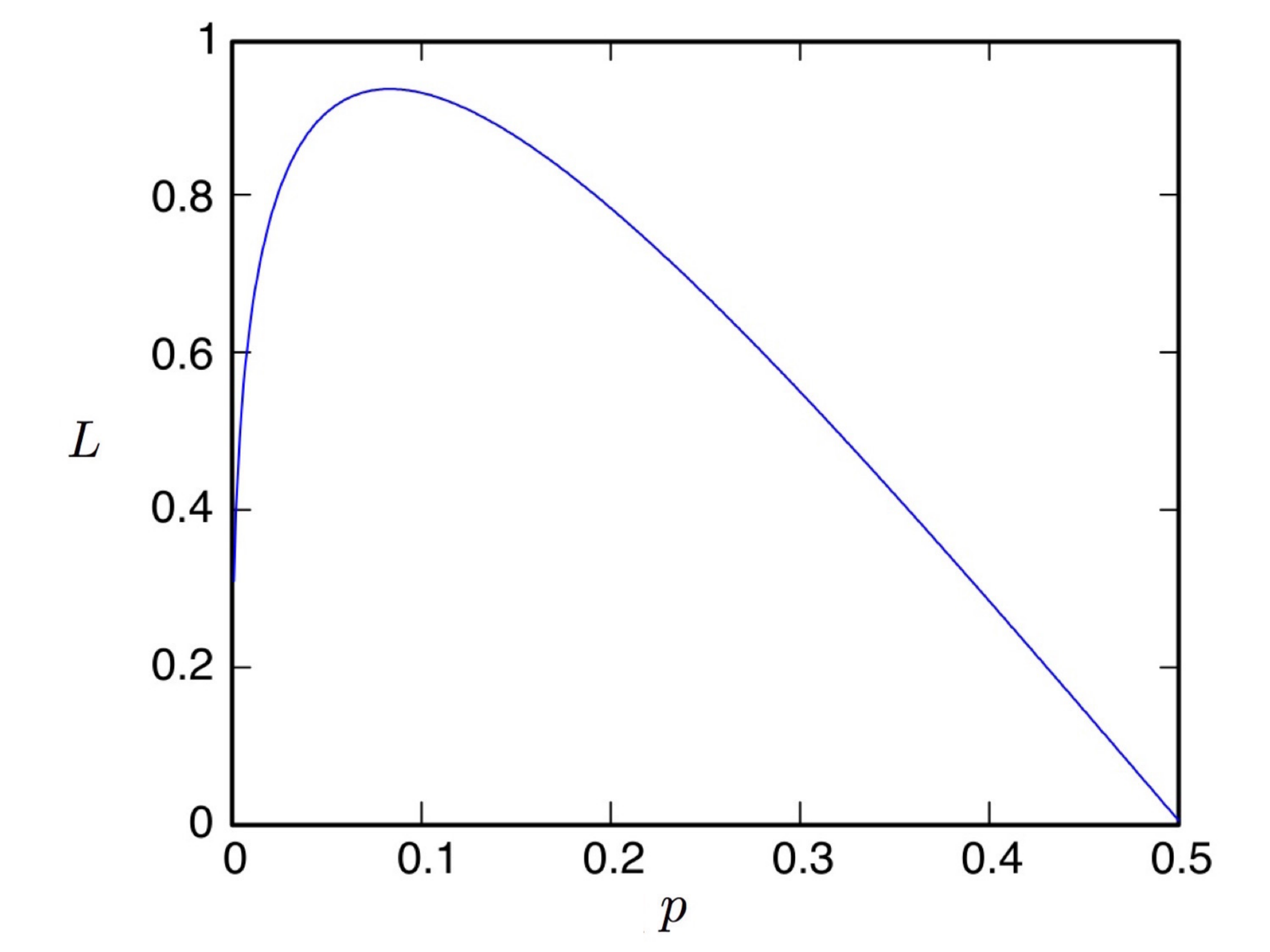}
\caption{The value of $L$ for the binary symmetric channel in Fig.~\ref{fig:BSC} as a function of $p$.} \label{fig:plot}
\end{figure}

Consider the binary symmetric channel in Fig.~\ref{fig:BSC}. Clearly, the only possible choice for $\tilde{P}$ in \eqref{eq:general} is $\tilde{P}(1)=1$. We thus obtain the value of $L$ as a function of $p$, which we plot in Fig.~\ref{fig:plot}.
Not surprisingly, when $p$ approaches $0.5$, $L$ approaches zero, as does the capacity of the channel. It is however interesting to notice that, when $p$ approaches zero, $L$ also approaches zero, even though the capacity of the channel approaches $1$ bit per use. This is because, when $p$ is very small, it is very easy to distinguish the two input symbols $0$ and $1$ at the receiver end. Hence the LPD criterion requires that the transmitter must use $1$ very sparsely, limiting the number of information bits it can send. The maximum of $L$ is approximately $0.94$ $\sqrt{\textnormal{nat}}$, achieved at $p=0.083$.

\begin{IEEEproof}[Proof of Theorem~\ref{thm:general}]
For every $n$, let
\begin{equation}
	\hat{P}_n \triangleq \argmax_{P_n} I(P_n, W)
\end{equation}
subject to
\begin{equation}\label{eq:DQn}
	D(Q_n\|Q_0) \le \frac{\delta}{n}.
\end{equation}
Using the same argument as for \eqref{eq:Pntoone}, we have
\begin{equation}
	\lim_{n\to\infty} \hat{P}_n(0) =1,
\end{equation}
hence $\hat{P}_n$ can be written as
\begin{equation}\label{eq:mixture}
	\hat{P}_n = (1-\mu_n) P_0 + \mu_n \tilde{P}_n
\end{equation}
where $P_0$ is the deterministic distribution with $P_0(0)=1$, $\tilde{P}_n$ is a distribution with $\tilde{P}_n(0)=0$, and $\mu_n$ is positive and tends to zero as $n$ tends to infinity. Fix $\tilde{P}_n$ and consider $\hat{P}_n$ given by \eqref{eq:mixture} as a function of $\mu_n$, then
\begin{equation}
	\left. \frac{\d I(\hat{P}_n,W)}{\d \mu_n} \right|_{\mu_n=0}= \sum_{x\in\set{X}} \tilde{P}_n(x) D( W(\cdot|x) \| Q_0),
\end{equation}
hence
\begin{equation}\label{eq:general33}
	I(\hat{P}_n, W) = \mu_n \sum_{x\in\set{X}} \tilde{P}_n(x) D( W(\cdot|x) \| Q_0) + o(\mu_n),
\end{equation}
where the term $o(\mu_n)$ tends to zero faster than $\mu_n$ as $n$ tends to infinity.

The output distribution resulting from feeding $\hat{P}_n$ given by \eqref{eq:mixture} into the channel $W$ is
\begin{equation}
	\hat{Q}_n = (1-\mu_n) Q_0 + \mu_n \tilde{Q}_n
\end{equation}
where $\tilde{Q}_n$ is the output distribution induced by input distribution $\tilde{P}_n$ through $W$. The relative entropy $D(\hat{Q}_n\| Q_0)$ is approximated by the Fisher Information \cite{kullback59} with respect to parameter $\mu_n$:
\begin{equation}\label{eq:fisher36}
	D(\hat{Q}_n\| Q_0) = \frac{\mu_n^2}{2} \sum_{y\in\set{Y}} \frac{\big(\tilde{Q}_n(y) - Q_0(y)\big)^2}{Q_0(y)} + o(\mu_n^2),
\end{equation}
where the term $o(\mu_n^2)$ tends to zero faster than $\mu_n^2$  as $n$ tends to infinity.
By \eqref{eq:DQn} and \eqref{eq:fisher36}, $\mu_n$ should have the form
\begin{equation}\label{eq:general36}
	\mu_n = \sqrt{\frac{\delta}{n}} \cdot \frac{1}{\sqrt{\displaystyle \frac{1}{2}\sum_{y\in\set{Y}} \frac{\big(\tilde{Q}_n(y) - Q_0(y)\big)^2}{Q_0(y)}}} + o\left(n^{-1/2}\right).
\end{equation}
Plugging \eqref{eq:general36}  into \eqref{eq:general33} yields
\begin{IEEEeqnarray}{rCl}
	 I(\hat{P}_n,W) & = & \sqrt{\frac{\delta}{n}}\cdot \frac{ \sum_{x\in\set{X}} \tilde{P}_n(x) D\left( \left.(W(\cdot|x) \right\| Q_0\right)}{\sqrt{\displaystyle \frac{1}{2}\sum_{y\in\set{Y}} \frac{\big(\tilde{Q}_n(y) - Q_0(y)\big)^2}{Q_0(y)}}} \nonumber\\
	& & {} + o\left(n^{-1/2}\right).\label{eq:general37}
\end{IEEEeqnarray}
When $n$ tends to infinity, $I(\hat{P}_n,W)$ is dominated by the first term on the right-hand side of \eqref{eq:general37}, hence $\tilde{P}_n$ should tend to the (not necessarily unique) distribution that maximizes this term.
Recalling Theorem~\ref{thm:IXY}, this completes the proof of Theorem~\ref{thm:general}.
\end{IEEEproof}

From the proof of Theorem~\ref{thm:general} it follows that the limit inferior in \eqref{eq:IXY} can be replaced by the limit, yielding a more convenient expression for $L$:

\begin{corollary}\label{cor:lim}
	For any DMC,
	\begin{equation}\label{eq:IXY2}
	L = \lim_{n\to\infty} \sqrt{\frac{n}{\delta}} \max_{P_n} I(P_n,W)
	\end{equation}
	where the maxima are subject to \eqref{eq:deltan}.
\end{corollary}
\begin{IEEEproof}
We only need to show that the limit in \eqref{eq:IXY2} exists. When input symbol $0$ is redundant, this limit exists and is infinity. When $0$ is not redundant, the proof of Theorem~\ref{thm:general} shows that this limit also exists and equals the right-hand side of \eqref{eq:general}.
\end{IEEEproof}

\section{A Simpler but Less General Expression for $L$}\label{sec:var}

In this section we consider channels that satisfy the following condition.
\begin{condition}\label{con:all}
There exists a capacity-achieving input distribution that uses all the input symbols.
\end{condition}

Note that Condition~\ref{con:all} implies that no input symbol is redundant; in particular, $0$ is not redundant.

We next give a simple upper bound on $L$ under Condition~\ref{con:all}. Later we provide an additional condition under which this bound is tight.


\begin{theorem}\label{thm:var}
	Consider a DMC that satisfies Condition~\ref{con:all}. Denote its capacity-achieving output distribution by $Q^*$, then
\begin{equation}\label{eq:var}
	L \le \sqrt{ 2 \,\mathsf{var}_{Q_0} \left(\log\frac{Q_0(Y)}{Q^*(Y)}\right)},
\end{equation}
where $\mathsf{var}_{Q_0}(\cdot)$ denotes the variance of a function of $Y$ where $Y$ has distribution $Q_0$.
\end{theorem}


The proof of Theorem~\ref{thm:var} utilizes the following lemma.

\begin{lemma}\label{lem:KT}
Let $Q^*$ denote the capacity-achieving output distribution for a DMC $W(\cdot|\cdot)$ of capacity $C$. Let $P'$ be any input distribution, and let $Q'$ denote the output distribution induced by $P'$ through $W$. Then
\begin{equation}\label{eq:lem}
	I(P',W) \le C - D(Q'\| Q^*),
\end{equation}
where equality holds if $\mathsf{supp}(P') \subseteq \mathsf{supp}(P^*)$ for some capacity-achieving input distribution $P^*$.
\end{lemma}

\begin{IEEEproof}
We have the following identity (see \cite{topsoe67}):
\begin{IEEEeqnarray}{rCl}
	I(P',W) & = & \sum_{x\in\set{X}} P'(x) D(W(\cdot|x)\| Q')\nonumber\\
	& = & \sum_{x\in\set{X}} P'(x) \E[W(\cdot|x)]{\log\frac{W(Y|x)}{Q'(Y)}}\nonumber\\
	& = & \sum_{x\in\set{X}} P'(x) \left(\E[W(\cdot|x)]{\log\frac{W(Y|x)}{Q^*(Y)}} \right.\nonumber\\
	& & ~~~~~~~\left. {}- \E[W(\cdot|x)]{\log\frac{Q'(Y)}{Q^*(Y)}}\right)\nonumber \\
	& = & \sum_{x\in\set{X}} P'(x) D(W(\cdot|x)\|Q^*) - D(Q'\|Q^*).\IEEEeqnarraynumspace \label{eq:topsoe}
\end{IEEEeqnarray}
By the Kuhn-Tucker conditions for channel capacity \cite{csiszarkorner81}, 
\begin{equation}
	D(W(\cdot|x)\| Q^*)) \le C
\end{equation}
where equality holds if $x\in\mathsf{supp}(P^*)$. We hence have
\begin{IEEEeqnarray}{rCl}
	C & = & \sum_{x\in\set{X}} P^*(x) D (W(\cdot|x)\|Q^*)\nonumber\\
	& \ge & \sum_{x\in\set{X}} P'(x) D (W(\cdot|x)\|Q^*), \label{eq:KKT}
\end{IEEEeqnarray}
where equality holds if $\mathsf{supp}(P') \subseteq \mathsf{supp}(P^*)$. Combining \eqref{eq:topsoe} and \eqref{eq:KKT} proves the lemma.
\end{IEEEproof}

\begin{IEEEproof}[Proof of Theorem~\ref{thm:var}]	
Since the channel satisfies Condition~\ref{con:all}, from Lemma~\ref{lem:KT} and Corollary~\ref{cor:lim} we have
\begin{equation}\label{eq:DQQ}
	L = \lim_{n\to\infty} \sqrt{\frac{n}{\delta}} \left(C - \min D(Q_n\|Q^*)\right),
\end{equation}
where the minimum is over  $Q_n\in\mathsf{conv}\{W(\cdot|x)\colon x\in\set{X}\}$ satisfying \eqref{eq:deltan}. To determine $L$, we need to find $Q_n$ that minimizes $D(Q_n\| Q_0)$ for a fixed $D(Q_n\|Q^*)$. To find an upper bound on $L$, we drop the condition $Q_n\in\mathsf{conv}\{W(\cdot|x)\colon x\in\set{X}\}$ to consider all distributions on $\set{Y}$. Then the minimum is well known to be achieved by a distribution from the exponential family connecting $Q_0$ and $Q^*$ \cite{csiszarmatus03}:
\begin{equation}\label{eq:Qlambda}
	Q_n (y) = \frac{Q_0(y)^{1-\lambda_n} Q^*(y)^{\lambda_n}}{\sum_{{y'}\in\set{Y}} Q_0(y')^{1-\lambda_n} Q^*(y')^{\lambda_n}},\quad y\in\set{Y}
\end{equation}
for some $\lambda_n\in[0,1]$. Indeed, if a distribution $Q_n$ minimizes $D(Q_n\| Q^*)$ for some fixed $D(Q_n\|Q_0)$, then it must minimize $$(1-\lambda_n)D(Q_n\|Q_0) + \lambda_n D(Q_n\|Q^*)$$ for some $\lambda_n\in[0,1]$. This sum can be written as
\begin{IEEEeqnarray}{rCl}
	\lefteqn{(1-\lambda_n)D(Q_n\|Q_0) + \lambda_n D(Q_n\|Q^*)}~~~~~\nonumber\\
	& = & D(Q_n\|R_n) - \log \sum_{y'\in\set{Y}} Q_0(y')^{1-\lambda_n} Q^*(y')^{\lambda_n},\IEEEeqnarraynumspace
\end{IEEEeqnarray}
where
\begin{equation}
	R_n (y) \triangleq \frac{Q_0(y)^{1-\lambda_n} Q^*(y)^{\lambda_n}}{\sum_{{y'}\in\set{Y}} Q_0(y')^{1-\lambda_n} Q^*(y')^{\lambda_n}},\quad y\in\set{Y}.
\end{equation}
Hence the best choice is $Q_n = R_n$. 

It remains to compute $D(Q_n\| Q_0)$ and $D(Q_n\| Q^*)$, where $Q_n$ is of the form \eqref{eq:Qlambda}, for large $n$. When $n$ is large, $Q_n$ must be close to $Q_0$ and hence $\lambda_n$ must be close to zero. In this case, $D(Q_n \| Q_0)$ is approximated by the Fisher Information~\cite{kullback59} with respect to parameter $\lambda_n$:
\begin{equation}
	D(Q_n\| Q_0) = \frac{\lambda_n^2}{2} \mathsf{var}_{Q_0} \left(\log\frac{Q_0(Y)}{Q^*(Y)}\right) + o(\lambda_n^2).
\end{equation}
This together with the requirement that $Q_n$ must satisfy \eqref{eq:deltan} implies that
\begin{equation}\label{eq:lambda}
	\lambda_n \le \sqrt{\frac{2\delta}{\displaystyle n \,\mathsf{var}_{Q_0} \left(\log\frac{Q_0(Y)}{Q^*(Y)}\right)}} + o(n^{-1/2}).
\end{equation}
Next we compute the derivative of $D(Q_n\|Q^*)$, with $Q_n$ given in \eqref{eq:Qlambda}, with respect to $\lambda_n$ evaluated at $\lambda_n=0$ to be
\begin{equation}
\left.\frac{\d D(Q_n\|Q^*)}{\d \lambda_n} \right|_{\lambda_n=0} = - \mathsf{var}_{Q_0}\left(\log\frac{Q_0(Y)}{Q^*(Y)}\right).
\end{equation}
By Condition~\ref{con:all}, there exists a capacity-achieving input distribution that uses $0$, so
\begin{equation}
	\lim_{\lambda_n\downarrow 0} D(Q_n\|Q^*) = D(Q_0\|Q^*) = C.
\end{equation}
Hence
\begin{equation}\label{eq:1stderivative}
	C - D(R_n\| Q^*) = \lambda_n \mathsf{var}_{Q_0} \left(\log\frac{Q_0(Y)}{Q^*(Y)}\right) + o(\lambda_n).
\end{equation}
Combining \eqref{eq:DQQ}, \eqref{eq:lambda}, and \eqref{eq:1stderivative} proves \eqref{eq:var}.
\end{IEEEproof}

The bound \eqref{eq:var} is tight for many channels, e.g., the binary symmetric channel of Example~\ref{ex:BSC}. We next provide a sufficient condition for \eqref{eq:var} to be tight.

Let $\mathbf{s}$ be the $|\set{Y}|$-dimensional vector given by
\begin{equation}
	s(y) = Q_0(y)\left(\log\frac{Q^*(y)}{Q_0(y)}+C\right),\quad y\in\set{Y}.
\end{equation}
Consider the following system of linear equations with unknowns $\alpha_x$, $x\in\set{X}\setminus\{0\}$:
\begin{equation}\label{eq:system}
	\sum_{x\in\set{X}\setminus\{0\}} \alpha_x \left( W(\cdot|x) - Q_0\right) = \vect{s}.
\end{equation}
Solving \eqref{eq:system} is a simple problem in linear algebra.

\begin{theorem}\label{thm:equalvar}
	Suppose Condition~\ref{con:all} is satisfied. If \eqref{eq:system} has a nonnegative solution, then \eqref{eq:var} holds with equality:
\begin{equation}\label{eq:equalvar}
	L = \sqrt{ 2 \,\mathsf{var}_{Q_0} \left(\log\frac{Q_0(Y)}{Q^*(Y)}\right)}.
\end{equation}	
\end{theorem}

The intuition behind Theorem~\ref{thm:equalvar} is the following: the vector $\vect{s}$ represents the tangent of the curve $Q_n(y)$ given by \eqref{eq:Qlambda} as a function of $\lambda_n$ at $\lambda_n=0$. That \eqref{eq:system} has a nonnegative solution means that $\vect{s}$ lies in the convex cone generated by $\{W(\cdot|x) - Q_0\colon x\in\set{X}\setminus\{0\}\}$. This further implies that, for small enough $\lambda_n$, $Q_n$ of the form given by \eqref{eq:system} is a valid output distribution, which, as can be seen in the proof of Theorem~\ref{thm:var}, guarantees \eqref{eq:var} to hold with equality. Along a different direction, we provide below a proof utilizing Theorem~\ref{thm:general}.

\begin{IEEEproof}[Proof of Theorem~\ref{thm:equalvar}]
We use Theorem~\ref{thm:general} to prove Theorem~\ref{thm:equalvar}. Let $\{\alpha_x\colon x\in\set{X}\setminus \{0\} \}$ be a nonnegative solution to \eqref{eq:system}, and let 
\begin{equation}
	A \triangleq \sum_{x\in\set{X}\setminus \{0\}} \alpha_x.
\end{equation}
Then the following constitutes a valid choice for $\tilde{P}$ in \eqref{eq:general}:
\begin{equation}\label{eq:tildeP}
\tilde{P}(x) = \frac{\alpha_x}{A},\quad x\in\set{X}\setminus \{0\}.
\end{equation}
The corresponding $\tilde{Q}$ is given by
\begin{IEEEeqnarray}{rCl}
	\tilde{Q} & = & \sum_{x\in\set{X}\setminus \{0\}} \frac{\alpha_x}{A} W(\cdot|x)\nonumber\\
	& = & Q_0+\frac{1}{A} \sum_{x\in\set{X}\setminus \{0\}} \alpha_x (W(\cdot|x) - Q_0)\nonumber\\
	& = & Q_0 + \frac{\vect{s}}{A}. \label{eq:tildeQ}
\end{IEEEeqnarray}
We evaluate \eqref{eq:general} for this choice of $\tilde{P}$ to obtain a lower bound on $L$. We first compute the denominator, using \eqref{eq:tildeQ}:
\begin{IEEEeqnarray}{rCl}
\lefteqn{\sqrt{\displaystyle \frac{1}{2}\sum_{y\in\set{Y}} \frac{\big(\tilde{Q}(y) - Q_0(y)\big)^2}{Q_0(y)}}}~~~~~~~~~~ \nonumber\\
& = & \sqrt{\frac{1}{2A^2} \sum_{y\in\set{Y}} \frac{s(y)^2}{Q_0(y)} }\nonumber\\
& = & \sqrt{\frac{1}{2A^2} \sum_{y\in\set{Y}} Q_0(y) \left(\log \frac{Q^* (y)}{Q_0(y)} + C \right)^2}\nonumber\\
& = & \sqrt{\frac{1}{2A^2} \,\mathsf{var}_{Q_0} \left(\log\frac{Q^*(Y)}{Q_0(Y)}\right)}. \label{eq:equalvardeno}
\end{IEEEeqnarray}
We next compute the numerator:
\begin{IEEEeqnarray}{rCl}
\lefteqn{\sum_{x\in\set{X}\setminus\{ 0\}} \tilde{P}(x) D\left( \left.(W(\cdot|x) \right\| Q_0\right)}~~~\nonumber\\
& = & \sum_{x\in\set{X}\setminus\{ 0\}} \tilde{P}(x) \sum_{y\in\set{Y}} W(y|x) \log \frac{W(y|x)}{Q^*(y)} \nonumber\\
& & {} + \sum_{x\in\set{X}\setminus\{ 0\}} \tilde{P}(x) \sum_{y\in\set{Y}} W(y|x) \log\frac{Q^*(y)}{Q_0(y)}\nonumber\\
& = & \sum_{x\in\set{X}\setminus\{ 0\}} \tilde{P}(x) \cdot C + \frac{1}{A}\sum_{\substack{x\in\set{X}\setminus\{ 0\}\\y\in\set{Y}}} \alpha_x W(y|x) \log\frac{Q^*(y)}{Q_0(y)}\nonumber\\
& = & C + \frac{1}{A} \sum_{y\in\set{Y}} \log\frac{Q^*(y)}{Q_0(y)} \sum_{x\in\set{X}\setminus\{ 0\}} \alpha_x W(y|x)\nonumber\\
& = & C + \frac{1}{A} \sum_{y\in\set{Y}} \log\frac{Q^*(y)}{Q_0(y)} \bigl(A Q_0(y) + s(y)\bigr) \label{eq:equalvar61}\\
& = & C - D(Q_0\| Q^*) + \frac{1}{A} \sum_{y\in\set{Y}} s(y) \log\frac{Q^*(y)}{Q_0(y)} \nonumber\\
& = & C - C + \frac{1}{A} \sum_{y\in\set{Y}} Q_0(y) \log\frac{Q^*(y)}{Q_0(y)} \left( \log\frac{Q^*(y)}{Q_0(y)} + C \right)\nonumber\\
& = & \frac{1}{A} \mathsf{var}_{Q_0} \left(\log\frac{Q^*(y)}{Q_0(y)}\right), \label{eq:equalvarnume}
\end{IEEEeqnarray}
where \eqref{eq:equalvar61} follows from \eqref{eq:system}. Combining Theorem~\ref{thm:general}, \eqref{eq:equalvardeno}, and \eqref{eq:equalvarnume} yields
\begin{equation}\label{eq:equalvar63}
	L \ge \sqrt{2\, \mathsf{var}_{Q_0} \left(\log\frac{Q^*(y)}{Q_0(y)}\right)}.
\end{equation}
Recalling Theorem~\ref{thm:var}, both \eqref{eq:var} and \eqref{eq:equalvar63} must hold with equality.
\end{IEEEproof}

\begin{example}
A $k$-ary uniform-error channel.
\end{example}

Consider a channel with $\set{X}=\set{Y}=\{0,1,\ldots,k-1\}$ and
\begin{equation}\label{eq:uniformerror}
	W(y|x) = \begin{cases} 1-p,& y=x\\ \displaystyle \frac{p}{k-1},&y\neq x\end{cases}
\end{equation}
where $p\in(0,1)$.
Clearly, its capacity-achieving output distribution $Q^*$ is uniform. It is easy to check that \eqref{eq:system} has solution
\begin{equation}
\alpha_x= \frac{p(1-p) \bigl(\log((k-1)(1-p)-\log p)\bigr)}{(k-1)(1-p)-p},\quad x\in\set{X}\setminus \{0\}
\end{equation}
which is nonnegative. We can hence use Theorem~\ref{thm:equalvar} to obtain
\begin{equation}
	L = \sqrt{2 v(k,p)}
\end{equation}
where
\begin{IEEEeqnarray}{rCl}
	v(k,p) & = & (1-p)\left( \log\frac{1}{1-p}\right)^2 + p \left(\log\frac{k-1}{p} \right)^2\nonumber\\
	& & {}- \left( (1-p)\log\frac{1}{1-p}+p\log\frac{k-1}{p}\right)^2.
\end{IEEEeqnarray}

While one might speculate that \eqref{eq:equalvar} holds, for example, for all symmetric channels, this is, perhaps surprisingly, not the case. The following example demonstrates this.

\begin{example}\label{ex:ternary}
	A ternary symmetric channel.
\end{example}

\begin{figure}[tbp]
\center
\includegraphics[width=0.5\textwidth]{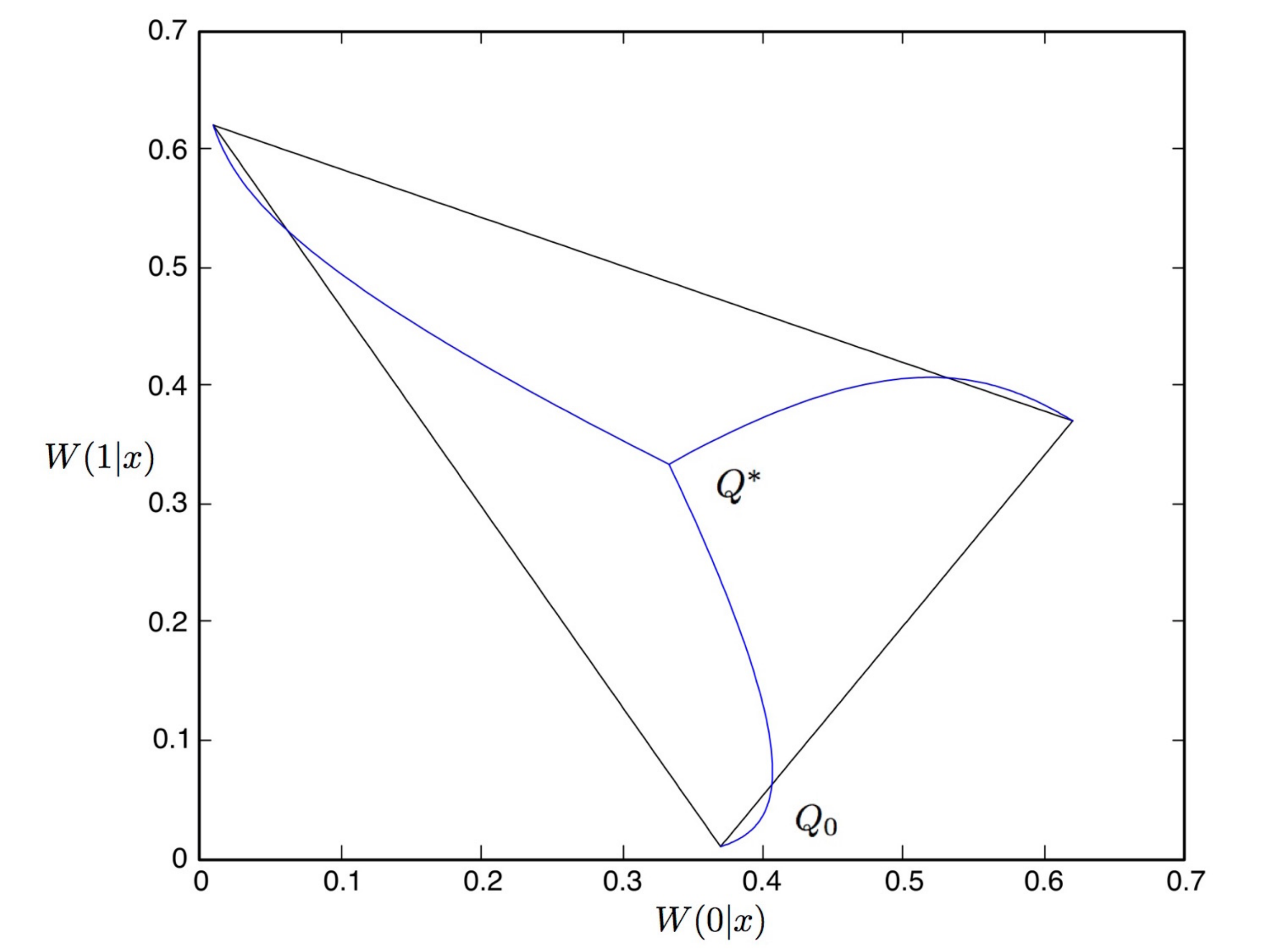}
\caption{The ternary symmetric channel in Example~\ref{ex:ternary}. The black triangle depicts the set of possible output distributions. The blue curves are the exponential families connecting the conditional output distributions and the capacity-achieving output distribution $Q^*$. The exponential family connecting $Q_0$ and $Q^*$ (as the other two exponential families) has a part that lies outside the black triangle, which is why \eqref{eq:equalvar} does \emph{not} hold for this channel.} \label{fig:exponential}
\end{figure}

Consider a ternary symmetric channel where $\set{X}=\set{Y}=\{0,1,2\}$ and
\begin{subequations}
\begin{IEEEeqnarray}{rCl}
W(\cdot| 0) & = & [ 0.37~~0.01~~0.62 ]\\
W(\cdot|1) & = & [ 0.62~~0.37~~0.01 ]\\
W(\cdot|2) & = & [ 0.01~~0.62~~0.37].
\end{IEEEeqnarray}
\end{subequations}
The right-hand side of \eqref{eq:equalvar} yields $0.66$ for this channel, but one can check that, in fact, $L=0.62$. This is because, as Fig.~\ref{fig:exponential} shows, the exponential family connecting $Q_0$ and $Q^*$ in the neighborhood of $Q_0$ does not lie in the set of possible output distributions $\mathsf{conv}\{W(\cdot|x)\colon x\in\set{X}\}$, or, roughly equivalently, $\vect{s}$ does not lie in the convex cone generated by $\{W(\cdot|x) - Q_0\colon x\in\set{X}\setminus\{0\}\}$.

\section{AWGN Channels}\label{sec:AWGN}

Consider an AWGN channel described by
\begin{equation}\label{eq:AWGNmodel}
	Y = X+Z,
\end{equation}
where $X\in\Reals$ is the channel input, $Y\in\Reals$ is the channel output, and $Z\in\Reals$ has the zero-mean Gaussian distribution of variance $\sigma^2$, denoted $\Normal{0}{\sigma^2}$, and is independent of $X$. Let the ``off'' input symbol be $0$, so $Q_0$ is also $\Normal{0}{\sigma^2}$. The encoder and decoder generate a random code as in Section~\ref{sec:setup} subject to the LPD constraint \eqref{eq:LPD}, and $L$ is again defined as in \eqref{eq:defL}. Note that we do not impose any average- or peak-power constraint on the input, but imposing such constraints will not affect the value of $L$ due to the stronger LPD constraint \eqref{eq:LPD}.\footnote{The LPD constraint requires that the average input power tend to zero as $n$ tends to infinity, hence rendering any additional average-power constraint inactive. As for peak-power constraints, our choice of input distribution to achieve $L$ is zero-mean Gaussian with vanishing variance. The influence of cutting the tail of such a distribution to meet any peak-power constraint will vanish as $n$ tends to infinity.}

\begin{theorem}\label{thm:AWGN}
For an AWGN channel,
\begin{equation}\label{eq:L1}
	L = 1\ \sqrt{\textnormal{nat}}
\end{equation}
irrespectively of the noise power $\sigma^2$.
\end{theorem}

The proof of Theorem~\ref{thm:AWGN} is divided into the converse part and the achievability part, and is given below.

\subsection{Converse for Theorem~\ref{thm:AWGN}}

Examining the proof of Theorem~\ref{thm:IXY}, we see that its converse part is valid for the AWGN channel. Hence 
\begin{equation}\label{eq:IXYAWGN}
	L \le \max_{\{P_n\}} \varliminf_{n\to\infty} \sqrt{\frac{n}{\delta}} I(P_n,W)
\end{equation}
where the maximum is taken over sequences of joint distributions on $(X,Y)\in \Reals\times\Reals$ induced by input distribution $P_n$ via the channel law $W$ resulting from the relation \eqref{eq:AWGNmodel}, such that the marginal distributions $Q_n$ for $Y$ satisfy 
\begin{equation}\label{eq:AWGNLPD}
D(Q_n\|Q_0)\le \frac{\delta}{n}.
\end{equation}

Let the second moment of the distribution $P_n$ be denoted $\rho_n$. It is well known that the zero-mean Gaussian maximizes $I(P_n,W)$ among all distributions of the same second moment (see, e.g., \cite{coverthomas91}), so
\begin{equation}\label{eq:AWGN65}
	I(P_n,W) \le \frac{1}{2}\log \left(1+\frac{\rho_n}{\sigma^2}\right).
\end{equation}
Because $X$ and $Z$ are independent, the second moment of the distribution $Q_n$ is $\rho_n+\sigma^2$, yielding
\begin{IEEEeqnarray*}{rCl}
D(Q_n\|Q_0) & = & -h(Q_n) + \E[Q_n]{\log\frac{1}{Q_0(Y)}}\\
	& = & -h(Q_n) + \E[Q_n]{ \log \left( \sqrt{2\pi\sigma^2} \,e^{\frac{Y^2}{2\sigma^2}}\right)}\\
	& = & -h(Q_n) + \frac{1}{2} \log \left(2\pi \sigma^2\right) + \E[Q_n]{\frac{Y^2}{2\sigma^2}}\\
	& = & -h(Q_n) +\frac{1}{2} \log \left(2\pi \sigma^2\right) + \frac{\rho_n+\sigma^2}{2\sigma^2}\cdot \\
	& \ge & -\frac{1}{2} \log \left(2\pi e (\rho_n+\sigma^2)\right) \\
	& & {} +\frac{1}{2} \log \left(2\pi \sigma^2\right) + \frac{\rho_n+\sigma^2}{2\sigma^2} \\
	& = & \frac{\rho_n}{2\sigma^2} -\frac{1}{2} \log\frac{\rho_n+\sigma^2}{\sigma^2}, \IEEEyesnumber\label{eq:AWGN66}
\end{IEEEeqnarray*}
where $h(\cdot)$ denotes the differential entropy, and where the inequality follows because the zero-mean Gaussian distribution maximizes differential entropy among all distributions of the same second moment. It follows from \eqref{eq:AWGN66} that, for $D(Q_n\|Q_0)$ to approach zero as $n$ tends to infinity, $\rho_n$ must tend to zero and
\begin{equation}
	D(Q_n\|Q_0) \ge \frac{\rho_n^2}{4\sigma^4} + o(\rho_n^2).
\end{equation}
Combined with \eqref{eq:AWGNLPD}, this implies
\begin{equation}
	\rho_n \le 2\sigma^2 \sqrt{\frac{\delta}{n}} + o(n^{-1/2}).
\end{equation}
Plugging this into \eqref{eq:AWGN65}  we obtain
\begin{IEEEeqnarray*}{rCl}
	I(P_n,W) & \le & \frac{1}{2}\log \left(1+\frac{\rho_n}{\sigma^2}\right)\\
	& \le & \frac{\rho_n}{2\sigma^2}\\
	& \le & \sqrt{\frac{\delta}{n}} + o(n^{-1/2}). \IEEEyesnumber\label{eq:AWGN68}
\end{IEEEeqnarray*}
Combining \eqref{eq:IXYAWGN} and \eqref{eq:AWGN68} yields
\begin{equation}
	L \le 1.
\end{equation}
This concludes the proof of the converse part of Theorem~\ref{thm:AWGN}.

\subsection{Achievability for Theorem~\ref{thm:AWGN}}

The achievability proof of Theorem~\ref{thm:IXY} relies on the finiteness of the input and output alphabets, therefore it is not applicable to the AWGN channel. Indeed, Theorem~\ref{thm:IXY} may not hold for a general continuous-alphabet channel. However, for the AWGN channel, we only need to prove an achievability result for Gaussian input distributions, which is much simpler than proving it for arbitrary input distributions.

For blocklength $n$, we randomly generate a codebook such that every codeword is independent of every other codeword, and is IID $\Normal{0}{\rho_n}$ with
\begin{equation}
	\rho_n \triangleq 2\sigma^2 \sqrt{\frac{\delta}{n}}.
\end{equation}

We first check that the LPD condition is met. Indeed, the output sequence is IID $\Normal{0}{\rho_n+\sigma^2}$, so
\begin{IEEEeqnarray*}{rCl}
D\left(Q^n \left\| Q_0^{\times n} \right.\right) & = & n D\left(\Normal{0}{\rho_n+\sigma^2} \| \Normal{0}{\sigma^2}\right)\\
	& = & n \left( \frac{\rho_n}{2\sigma^2} - \frac{1}{2} \log\frac{\rho_n+\sigma^2}{\sigma^2} \right)\\
	& \le & n \left(\frac{\rho_n}{2\sigma^2} - \frac{1}{2} \left(\frac{\rho_n}{\sigma^2} - \frac{\rho_n^2}{2\sigma^4}\right)\right)\\
	& = & \frac{n\rho_n^2}{4\sigma^4}\\
	& = & \frac{n}{4\sigma^4} \cdot \left(2\sigma^2 \sqrt{\frac{\delta}{n}}\right)^2\\
	& = & \delta, \IEEEyesnumber
\end{IEEEeqnarray*}
where for the inequality we use the fact
\begin{equation}\label{eq:logineq}
	\log (1+a) \ge a - \frac{a^2}{2}, \quad a\ge 0.
\end{equation}

We next look at the maximum number of nats that can be reliably transmitted with this code. Similar to the DMC case, we can show that the sequence $\{K_n\}$ is achievable if \eqref{eq:Pliminf} holds, except that now $Q_n$ and $W$ are density and conditional density, respectively. The ratio between $W$ and $Q_n^{\times n}$ in \eqref{eq:Pliminf} can be evaluated as
\begin{IEEEeqnarray*}{rCl}
	\lefteqn{\frac{W(y^n|x^n)}{Q_n^{\times n} (y^n)}}~~~~~~\\
	 & = & \frac{\displaystyle \prod_{i=1}^n \frac{1}{\sqrt{2\pi\sigma^2}} e^{-\frac{(y_i-x_i)^2}{2\sigma^2}}}{\displaystyle \prod_{i=1}^n \frac{1}{\sqrt{2\pi(\rho_n+\sigma^2)}} e^{-\frac{y_i^2}{2(\rho_n+\sigma^2)}}}\\
	& = & \left(\frac{\rho_n+\sigma^2}{\sigma^2}\right)^\frac{n}{2} \exp \left( \frac{\sum_{i=1}^n y_i^2}{2(\rho_n+\sigma^2)} - \frac{\sum_{i=1}^n z_i^2}{2\sigma^2}\right).\IEEEyesnumber\IEEEeqnarraynumspace
\end{IEEEeqnarray*}
Hence 
\begin{IEEEeqnarray}{rCl}
\frac{1}{\sqrt{n}} \log \frac{W(Y^n|X^n)}{Q_n^{\times n} (Y^n)}
	& = & \frac{\sqrt{n}}{2} \log \left(\frac{\rho_n + \sigma^2}{\sigma^2} \right) \nonumber\\
	& & {} + \frac{1}{\sqrt{n}} \left( \frac{\sum_{i=1}^n Y_i^2}{2(\rho_n+\sigma^2)} - \frac{\sum_{i=1}^n Z_i^2}{2\sigma^2}\right). \nonumber \\ \,  \label{eq:yes73}
\end{IEEEeqnarray}
The mean of \eqref{eq:yes73} satisfies
\begin{IEEEeqnarray*}{rCl}
	\lefteqn{ \E{\frac{1}{\sqrt{n}} \log \frac{W(Y^n|X^n)}{Q_n^{\times n} (Y^n)}}}~~~~\\
	& = & \frac{\sqrt{n}}{2} \log \left(\frac{\rho_n+\sigma^2}{\sigma^2}\right) \\
	& & {} + \frac{1}{\sqrt{n}} \left( \frac{\sum_{i=1}^n \E{Y_i^2}}{2(\rho_n+\sigma^2)} - \frac{\sum_{i=1}^n \E{Z_i^2}}{2\sigma^2}\right)\\
	& = & \frac{\sqrt{n}}{2} \log \left(\frac{\rho_n+\sigma^2}{\sigma^2}\right) + 0 \\
	& \ge & \frac{\sqrt{n}}{2} \left( \frac{\rho_n}{\sigma^2} - \frac{\rho_n^2}{2\sigma^4} \right)\\
	& = & \sqrt{\delta} - \frac{\delta}{\sqrt{n}},\IEEEyesnumber \label{eq:AWGN75}
\end{IEEEeqnarray*}
where we again use \eqref{eq:logineq}.
By \eqref{eq:AWGN75} we know that 
\begin{equation}
	\varliminf_{n\to\infty} \E{\frac{1}{\sqrt{n}} \log \frac{W(Y^n|X^n)}{Q_n^{\times n} (Y^n)}} \ge \sqrt{\delta}.
\end{equation}
It remains to show that 
\begin{equation}\label{eq:AWGNvartozero}
\lim_{n\to\infty} \mathsf{var} \left(\frac{1}{\sqrt{n}} \log \frac{W(Y^n|X^n)}{Q_n^{\times n} (Y^n)}  \right) = 0. 
\end{equation}
Then, by Chebyshev's inequality, we can establish
\begin{equation}\label{eq:AWGNliminf}
P-\liminf_{n\to\infty} \frac{1}{\sqrt{n}} \log \frac{W(Y^n|X^n)}{Q_n^{\times n} (Y^n)} \ge \sqrt{\delta}
\end{equation}
and hence
\begin{equation}\label{eq:AWGNachievability}
	L \ge 1.
\end{equation}
Using \eqref{eq:yes73}, the variance in \eqref{eq:AWGNvartozero} can be computed as:
\begin{IEEEeqnarray*}{rCl}
\lefteqn{ \mathsf{var} \left(\frac{1}{\sqrt{n}} \log \frac{W(Y^n|X^n)}{Q_n^{\times n} (Y^n)}  \right)}~~~\\
	& = & \mathsf{var}\left(\frac{1}{\sqrt{n}} \left( \frac{\sum_{i=1}^n Y_i^2}{2(\rho_n+\sigma^2)} - \frac{\sum_{i=1}^n Z_i^2}{2\sigma^2}\right) \right)\\
	& = & \frac{1}{n} \sum_{i=1}^n \mathsf{var} \left( \frac{Y_i^2}{2(\rho_n+\sigma^2)} - \frac{Z_i^2}{2\sigma^2} \right)\\
	& = & \mathsf{var} \left(\frac{Y^2}{2(\rho_n+\sigma^2)}-\frac{Z^2}{2\sigma^2} \right)\\
	& = & \E{\left(\frac{Y^2}{2(\rho_n+\sigma^2)}-\frac{Z^2}{2\sigma^2} \right)^2}\\
	& = & \frac{1}{4(\rho_n+\sigma^2)^2} \E{ \left(X^2+2XZ - \frac{\rho_n}{\sigma^2} Z^2 \right)^2}\\
	& \le & \frac{1}{4\sigma^4}\E{ \left(X^2+2XZ - \frac{\rho_n}{\sigma^2} Z^2 \right)^2}.\IEEEyesnumber\label{eq:AWGN81}
\end{IEEEeqnarray*}
After expanding the square inside the expectation in \eqref{eq:AWGN81}, one can verify that the expectation of every summand tends to zero as $n$ tends to infinity, establishing \eqref{eq:AWGNvartozero}, and hence \eqref{eq:AWGNliminf} and \eqref{eq:AWGNachievability}, proving the achievability part of Theorem~\ref{thm:AWGN}.

\section{Concluding Remarks}\label{sec:conclusion}

A DMC in practice often represents discretization of a continuous-alphabet channel. For example, Figs.~\ref{fig:BSC0} and~\ref{fig:BSC} can result from two different discretizations of the same AWGN channel. In this sense, our results suggest that the optimal discretization may depend heavily on whether there is an LPD requirement or not.

In practice, LPD communication systems of positive data rates often can be implemented even when the channel model does not seem to allow positive rates. Indeed, in such applications, the concern is often not that the transmitted signal should be sufficiently weak, but rather that it should have a wide spectrum and resemble white noise \cite{simon94}. We believe that one of the reasons why such systems may work is that realistic channels often have memory. For example, on a channel whose noise level varies with a coherence time that is longer than the length of a codeword, the transmitter and the receiver can use the adversary's ignorance of the actual noise level to communicate without being detected. One way to formulate this scenario is to assume that the channel has an unknown parameter that is fixed. This is discussed for the binary symmetric channel in \cite{chebakshichanjaggi14}. Further addressing this scenario is part of ongoing research.

\section*{Acknowledgements}

The authors thank Boulat Bash and Matthieu Bloch for helpful comments.


\bibliographystyle{hieeetr}           
\bibliography{/Volumes/Data/wang/Library/texmf/tex/bibtex/header_short,/Volumes/Data/wang/Library/texmf/tex/bibtex/bibliofile}

\end{document}